\documentstyle[12pt]{article}
\newcommand{\be}{\begin{equation}}
\newcommand{\ee}{\end{equation}}
\newcommand{\bea}{\begin{eqnarray}}
\newcommand{\ena}{\end{eqnarray}}
\newcommand{\beano}{\begin{eqnarray*}}
\newcommand{\enano}{\end{eqnarray*}}
\newcommand{\sect}[1]{\setcounter{equation}{0}\section{#1}}
\newcommand{\vs}[1]{\rule[- #1 mm]{0mm}{#1 mm}}
\newcommand{\hs}[1]{\hspace{#1 mm}}

\newcommand{\cc}{\mbox{$\cal{C}$}}

\newcommand{{\cg}}{\mbox{$\cal{G}$}}
\newcommand{\ch}{\mbox{$\cal{H}$}}

\newcommand{\cw}{\mbox{$\cal{W}$}}

\newcommand{\bal}{{\overline{\alpha}}}
\newcommand{\bbet}{{\overline{\beta}}}
\newcommand{\bgam}{{\overline{\gamma}}}

\newcommand{\prt}{\partial}
\newcommand{\eps}{\epsilon}

\newcommand{\wh}[1]{\widehat{#1}}
\newcommand{\wt}[1]{\widetilde{#1}}
\newcommand{\mb}[1]{\hs{5}\mbox{#1}\hs{5}}

\newcommand{\su}{\mbox{${S\ell(2)}$}}
\newcommand{\osp}{\mbox{${OSp(1|2)}$}}

\newtheorem{prop}{Property}

\newcommand{\NP}[1]{Nucl.\ Phys.\ {\bf #1}}
\newcommand{\PL}[1]{Phys.\ Lett.\ {\bf #1}}

\newcommand{\CMP}[1]{Comm.\ Math.\ Phys.\ {\bf #1}}

\begin{document}
\renewcommand{\thefootnote}{\fnsymbol{footnote}}
\newpage
\setcounter{page}{0}
\pagestyle{empty}

\vs{30}

\begin{center}

{\LARGE {\bf Gauged $\cw$ algebras}}\\[1cm]

\vs{10}

{\large F. Delduc$^1$, L. Frappat$^2$, E. Ragoucy$^{2}$, and
P. Sorba$^{1,2}$}

{\em Laboratoire de Physique Th\'eorique }
{\small E}N{\large S}{\Large L}{\large A}P{\small P}
\footnote{URA 14-36 du CNRS, associ\'ee \`a l'Ecole Normale
Sup\'erieure de
Lyon, et au Laboratoire d'Annecy-le-Vieux de Physique des Particules
(IN2P3-CNRS).

\noindent
$^1$ Groupe de Lyon: ENS Lyon, 46 all\'ee d'Italie, F-69364 Lyon
Cedex 07,France.

\noindent
$^2$ Groupe d'Annecy: LAPP, Chemin de Bellevue BP 110, F-74941
Annecy-le-Vieux Cedex, France.
}\\

\end{center}
\vs{20}

\centerline{ {\bf Abstract}}

\indent

We perform an Hamiltonian reduction on a classical \cw(\cg, \ch)
algebra, and prove that we get
another \cw(\cg, \ch$'$) algebra, with $\ch\subset\ch'$.
In the case $\cg=S\ell(n)$, the existence of a suitable gauge, called
Generalized Horizontal Gauge, allows to relate in this way two
\cw-algebras as
soon as their corresponding \ch-algebras are related by inclusion.

\vfill
\rightline{{\small E}N{\large S}{\Large L}{\large A}P{\small
P}-AL-460/94}
\rightline{March 1994}

\newpage
\pagestyle{plain}
\renewcommand{\thefootnote}{\arabic{footnote}}

\sect{Introduction}

\indent

One knows today how to associate a \cw-algebra
to each couple formed by a
simple Lie (super)algebra \cg\ and one of its \su\ embeddings
\cite{ORaf,Bais,FRS,ferm}.
Each algebra obtained in this way
appears as the symmetry algebra of a Toda theory \cite{Sav},
which itself can be seen, in a Lagrangian formalism, as a gauged
Wess-Zumino-Witten (WZW) model \cite{ORaf,BS}.
The supersymmetric generalization of
this framework, leading to a classification of \cw-superalgebras
associated to couples $\{$simple Lie superalgebras
$-$ \osp  embeddings$\}$,
is also known \cite{DRS,FRS}.
First studied in the classical case, the method has
also been extended to the quantum case \cite{BoTj,Sev,JOM}.

Considering the large class of \cw-algebras
and superalgebras so obtained, there
is no doubt that several properties of simple Lie (super)algebras could be used
to relate among them different \cw-(super)algebras.
Simplifying
their classification, this will shed some light on their structure and
motivate applications. For example, the technics of
folding, allowing to relate, via symmetry of their Dynkin diagrams, simply
laced algebras with non simply laced ones, has been adapted to the case of
\cw-(super)algebras \cite{fold}.
Therefore, it is possible, at least at the classical level,
to deduce the \cw\ families associated to the orthogonal
or symplectic (orthosymplectic in the
supersymmetric case) series from those relative to the unitary ones.

In this paper, connections among \cw-algebras relative to
the same \cg\ but different
\su\ embeddings will also be established. At this
point, let us recall that any \su\
embedding of a simple Lie algebra \cg\ can be seen,
up to some exceptions in the
$D_n$ and $E_n$ cases, as the principal \su\ of at
least one regular subalgebra \ch\
of \cg. We use \cw(\cg, \ch)
to denote the corresponding \cw-algebra. As we will see in the
following, at least for $\cg=S\ell(n)$, the algebra \cw(\cg, \ch$'$)=\cw$'$
can be constructed from \cw(\cg, \ch)=\cw, with $\ch\subset\ch'$,
by imposing a set
of constraints $\cc(\cw\rightarrow \cw')$ on the $W$ generators
and performing gauge transformations associated to these constraints.
Thus we
perform an Hamiltonian reduction in the same way that
the \cw-algebra itself is
obtained by constraining some \cg\ Kac-Moody currents
and determining quantities
which are invariant under the gauge group ($G$-subgroup)
relative to the corresponding set
of constraints $\cc(\cg\rightarrow\cw)$.
It can appear natural to call {\it secondary reduction}
such a procedure which allows to build the algebra \cw$'$ from the
algebra \cw\ by
gauging a part of this last one.

\indent

We start this letter by a rapid summary on Dirac constraints theory
applied to the determination of \cw-algebras (section \ref{sec2}).
Then, we show in section \ref{sec3} that the reduction $\cg
\rightarrow \cw(\cg,\ch')\equiv\cw'$ can be decomposed into two steps: $\cg
\rightarrow \cw(\cg,\ch)\equiv\cw$ followed by $\cw(\cg,\ch)
\rightarrow \cw(\cg,\ch')$ with $\ch\subset\ch'$, as soon as the set of
all constraints $\cc(\cg\rightarrow\cw)$ is included into the set
$\cc(\cg\rightarrow\cw')$. For \cg=$Sl(n)$, we introduce the
Generalized Horizontal Gauge (section \ref{sec4}) which insures that
\cw(\cg, \ch) can be related to \cw(\cg, \ch$'$) as soon as
$\ch\subset\ch'$. We illustrate our results in the $S\ell(4)$ case
(section \ref{sec5}) before concluding.

\newpage
\sect{Constraints and W algebras: a reminder\label{sec2}}

\indent

The construction of \cw-algebras via Hamiltonian reduction has been extensively
discussed in several papers \cite{ORaf,BS};
therefore, we will limit ourselves
introducing the main tools necessary for our purpose.

We start with the Lie algebra \cg\ of the non compact group G. Let
$t^a$, $a=1,\dots,dim\cg$, be a matrix
realization of a \cg-basis; we denote by $J=J_a\ t^a$
the matrix current we want to constrain.
The $J_a$ can be seen either as the right $J_+(x_+)$ or as the left
$J_-(x_-)$ conserved
currents, i.e. $\prt_\pm(J_\mp)=0$ of the WZW action $S(g)$ with
$g\in G$.

Let $M_0,M_\pm$ be the generators of the \su\
subalgebra of \cg\ chosen for the reduction. There exists
at least one regular \cg\ subalgebra (we disgard the exceptions arising in the
$D_n$ and $E_n$ cases) \ch\ for which this \su\  is principal. The gradation on
\cg\ induced
by $M_0$ leads to the decomposition:
\be
\cg=\oplus_k\cg_k =\cg_-\oplus\cg_0\oplus\cg_+
\ee
with obviously $M_\pm\in\cg_{\pm1}$.

Now we demand the $\cg_-$ part of $J$ to satisfy the following
conditions:
\be
\left. J\right|_{\cg_-}= M_- \label{Jcont}
\ee
which in particular implies for the $J$-components associated to elements of
$\cg_{<-1}$, if they exist, to be zero.

The conditions (\ref{Jcont}) are first class
constraints\footnote{In the case of half-integral
grading, some of the constraints we
impose may be second class, but this will not change the reasoning and
conclusion hereafter.} $C^{(1)}_A=0$ with $A=1,\dots,dim\cg_-$,
and generate through the Poisson Brackets (PB)
a gauge group with Lie algebra $\cg_+$.

The PB algebra of the current components is given by:
\be
\{ J_a(z), J_b(w)\}_{PB}= {f_{ab}}^c\ J_c(w)\ \delta(z-w)
+k\ \delta_{ab}\ \delta'(z-w)
\ee
where ${f_{ab}}^c$ are the structure constants for a given
basis of \cg.
Moreover, the gauge transformations associated to the
first class constraints $C^{(1)}_A$ act on $J_a$ as
\bea
J_a(w) &\rightarrow& J_a(w) +\int dz\ \eps^A(z) \{C^{(1)}_A(z),
J_a(w)\}_{PB}+ \nonumber\\
&&\mb{} +\frac{1}{2!} \int dy\int dz\ \eps^A(z)\eps^B(y)
\{C^{(1)}_B(y),\ \{ C^{(1)}_A(z),\ J_a(w)\}\} +\dots \label{eq.CJC}
\ena
where $\eps^A(x)$ is the parameter associated to $C^{(1)}_A(x)$.
As usual in Dirac constraints
theory \cite{Dirac}, the condition $C^{(1)}_A(x)=0$
has to be imposed after working out a $PB$, that is
why the name {\it weak equation} is often used.

The bracket of two $W$ elements can be computed
{\it \`a la Dirac}. From this point of view,
we have to take into account the second class
constraints determined by the gauge fixing. Let us denote $\{C_a(z)\}$ the
set of all
(first and second class) constraints: its dimension is even and equal
to $2p$, where $p$ is also
the dimension of the subgroup of \cg\ which has been gauged.
We also introduce:
\be
C_{AB}(x,y)= \{C_A(x), C_B(y)\}_{PB}
\ee
and its inverse $C^{AB}(x,y)$ such that:
\be
\int dy\ C_{AB}(x,y)C^{BD}(y,t)= \delta_A^D\ \delta(x-t) \label{C-1}
\ee
The (Dirac) brackets for the \cw(\cg, \ch) algebra read:
\be
\{J_i(z), J_j(w)\}_D= \{J_i(z), J_j(w)\}_{PB}-
\int\ dx\ dy\ \{J_i(z), C_A(x)\}_{PB}C^{AB}(x,y) \{ C_B(y), J_j(w)\}_{PB}
\label{Dirac}
\ee
In the following we will drop the $z$ dependence, and forget the
index $PB$ (while keeping the $D$ one). Thus, we will write
\bea
C_{AB}C^{BD} &=& \delta_A^D \label{eq.211} \\
\{J_i, J_j\}_D &=& \{J_i, J_j\} -\{J_i, C_A\}C^{AB}\{C_B,J_j\}
\ena
instead of (\ref{C-1}) and (\ref{Dirac}).

\sect{Secondary reduction\label{sec3}}

\indent

The following property will be specially usefull for our purpose:

\begin{prop}
Let \cw(\cg, \ch$'$) obtained from \cg\
by imposing the set $\{ C_A\}$ of first and
second class constraints. Let \cw(\cg, \ch) a second \cw-algebra with
$\ch\subset\ch'$ and such that the corresponding set
of constraints $\{ C_\alpha\}$ is a subset of $\{ C_A\}$.
Then, by
imposing on \cw(\cg, \ch) the constraints $\{ C_\bal\}$
such that  $\{ C_A\}=\{ C_\alpha\}\cup \{ C_{\bar{\alpha}}\}$,
we recover the algebra \cw(\cg, \ch$'$). \label{prop1}
\end{prop}
This property can be sumarized in the diagram:

\be
\begin{array}{l}
\begin{array}{ccccccccc}
\cg &\rightarrow &\rightarrow &\rightarrow &\{ C_A\}
&\rightarrow &\rightarrow &\rightarrow
& \cw(\cg,\ch') \\
 & \searrow & & & & & &\nearrow & \\
 & & \{ C_\alpha\} & & & &\{ C_{\bar\alpha}\} & & \\
 & & & \searrow & &\nearrow & & & \\
 & & & & \cw(\cg, \ch) & & & &
\end{array}\\ \\
\mb{with} \{ C_A\}=\{C_\alpha\} \cup \{ C_{\bar{\alpha}}\}
\end{array}
\ee

This property is a general feature of the Dirac Brackets (see e.g.
\cite{sunder}). We will show hereafter that it can be applied to the
case of \cw-algebras.

Let us remark that the number of free fields in
\cw(\cg, \ch$'$) and in the algebra
obtained by reducing \cw(\cg, \ch) via $\{ C_{\bar\alpha}\}$-constraints
is the same, since the
number of constraints imposed in both paths is identical.
Let us denote by $\{\cdot,\cdot\}_{D}$
the Dirac bracket associated to the reduction
$\cg\rightarrow\cw(\cg,\ch')$; $\{\cdot,\cdot\}_{D1}$ those associated
to $\cg\rightarrow\cw(\cg,\ch)$ and $\{\cdot,\cdot\}_{D2}$
those constructed from \cw(\cg, \ch) by imposing the constraints $\{
C_{\bar\alpha}\}$.

To carry on the proof, let us add the following notations:
in \cw(\cg, \ch) the free fields will be
$\{W_i,W_{\overline{\alpha}}\}\equiv\{W_{a}\}$,
and Dirac brackets will read:
\be
\{W_{a},W_{b}\}_{D1}= \{W_{a},W_{b}\}- \{W_{a},C_\alpha\}
\Gamma^{\alpha\beta}\{C_\beta,W_{b}\} \label{eq.D1}
\ee
where we have defined
\be
\Gamma_{\alpha\beta}= \{C_\alpha, C_\beta\}=C_{\alpha\beta}
\ee
and
\be
\Gamma^{\alpha\beta} \mb{such that} \Gamma_{\alpha\beta}\Gamma^{\beta\gamma}=
\delta_\alpha^\gamma
\ee

\indent

Now we want to constrain the \cw(\cg, \ch) elements which correspond to the
fields
$W_{\overline{\alpha}}$.
This means that we introduce new brackets on \cw(\cg, \ch):
\be
\{W_i, W_j\}_{D2}= \{W_i, W_j\}_{D1}- \{W_i, C_\bal\}_{D1}
\Delta^{\bal\bbet} \{C_\bbet, W_j\}_{D1} \label{eq.D2}
\ee
with
\be
\Delta_{\bal\bbet}= \{W_{\bal}, W_\bbet\}_{D1} \mb{;}
W_\bal\equiv C_\bal \mb{and}
\Delta_{\bal\bbet} \Delta^{\bbet\bgam}= \delta_\bal^\bgam
\ee

The property \ref{prop1} will be proved if one shows that
$\{ W_i,W_j\}_{D}=\{ W_i,W_j\}_{D2}$
for any couple of elements $(W_i,W_j)$ in \cw(\cg, \ch$'$) $-$or
in the $\{ C_{\bar\alpha}\}$-constrained \cw(\cg, \ch)
algebra.

Let us develop (\ref{eq.D2}) using (\ref{eq.D2})
\beano
\{W_i, W_j\}_{D2} &=& \{W_i, W_j\} -\{W_i, C_\alpha\}
\Gamma^{\alpha\beta} \{C_\beta, W_j\}+ \\
&& -\left( \{W_i, C_{\bal}\}
-\{W_i, C_\alpha\} \Gamma^{\alpha\beta} \{C_\beta, C_\bal\} \right)
\Delta^{\bal\bbet} \left( \{C_\bbet, W_j\} -\{C_\bbet, C_\gamma\}
\Gamma^{\gamma\sigma} \{C_\sigma, W_j\}\right)\\
&=& \{W_i, W_j\} -\{W_i, C_\alpha\} \left( \Gamma^{\alpha\beta}-
\Gamma^{\alpha\gamma}C_{\gamma\bal}\Delta^{\bal\bbet}
C_{\bbet\sigma}\Gamma^{\sigma\beta}
\right) \{C_{\beta},W_j\}+ \\
&&+\{W_i, C_\bal\} \Delta^{\bal\bbet}C_{\bbet\beta}C^{\beta\alpha}
\{C_\alpha, W_j\}
+\{W_i, C_\alpha\} \Gamma^{\alpha\beta}C_{\beta\bbet}\Delta^{\bbet\bal}
\{C_\bbet, W_j\}+ \\
&& -\{W_i, C_\bal\} \Delta^{\bal\bbet} \{C_\bbet, W_j\}
\enano
This last expression can, hopefully, be simplified owing to the following
identities:
\bea
C^{\alpha\beta} &=& \Gamma^{\alpha\beta}
-\Gamma^{\alpha\gamma}C_{\gamma\bgam}C^{\bgam\beta} \\
C^{\alpha\bal} &=&
-\Gamma^{\alpha\gamma}C_{\gamma\bgam}C^{\bgam\bal} \\
C^{\bal\alpha} &=&
-C^{\bal\bgam}C_{\bgam\gamma}\Gamma^{\gamma\alpha}
\ena
which are obtained by multiplying by $\Gamma^{\alpha\beta}$
both sides (\ref{eq.211}) with the couple (A,D)
being formed with one or two non bared greek indices. We will
also need the relations
\be
\Delta_{\bal\bbet}= C_{\bal\bbet}-
C_{\bal\alpha}\Gamma^{\alpha\beta}C_{\beta\bbet}
\ee
and
\be
\Delta^{\bal\bbet}=C^{\bal\bbet}
\ee
Then,
\beano
\{W_i, W_j\}_{D2} &=& \{W_i, W_j\} -\{W_i, C_\alpha\}
C^{\alpha\beta} \{C_\beta, W_j\}
-\{W_i, C_\bal\} C^{\bal\beta} \{C_\beta, W_j\}+ \\
&& -\{W_i, C_\alpha\} C^{\alpha\bbet} \{C_\bbet, W_j\}
-\{W_i, C_\bal\} C^{\bal\bbet} \{C_\bbet, W_j\}\\
&=& \{W_i, W_j\} -\{W_i, C_A\} C^{AB} \{C_B, W_j\} \\
&=& \{W_i, W_j\}_D
\enano

The simplest example of such a secondary reduction is the gauging of the
algebra $W_3^{(2)}$ also called Bershadsky-Polyakov algebra in order to obtain
the
$W_3$ or Zamolodchikov algebra. This case has been studied in
\cite{mexico}. We
consider other examples based on $S\ell(4)$ in section
\ref{sec5}. Before that, let us
discuss the general case of $\cw[S\ell(n-1),\ \ch]$ algebras.

\sect{A generalized horizontal gauge for $\cw[S\ell({n-1}),\ \ch]$
algebras\label{sec4}}

\indent

As developed in section \ref{sec3}, a sufficient condition
for obtaining the \cw(\cg, \ch$'$) algebra
from the \cw(\cg, \ch) one is the inclusion
of the constraints relative to \cw(\cg, \ch)
in those relative to \cw(\cg, \ch$'$). For $\cg=S\ell(n)$, we
introduce a general gauge which
guarantees the gauge passing $\cw(\cg, \ch)\rightarrow\cw(\cg, \ch')$
as soon as $\ch\subset\ch'$.

\indent

One knows that in $S\ell(n)$ , any \su-subalgebra is principal
in a regular subalgebra which itself is of the form
\be
\ch=\oplus_{i=1}^k\ S\ell(n_i) \mb{with} \left(\sum_{i=1}^k n_i\right)
\leq n
\ee
Note that one may have $n_i=n_j$ while $i\neq j$.

Considering a
$n\times n$ matrix representing the general element of the algebra,
let us start by
fixing a block diagonal part made with each $S\ell(n_i)$
algebra of \ch. We complete it by an $G\ell(p)$ subalgebra, where
$p=n-\sum_{i=1}^k n_i$, and by $(k-1)$ $U(1)$
generators, commuting with each $S\ell(n_i)$ and $G\ell(p)$ part.
On each $S\ell(n_i)$ block, there will be 1
on each entry of the minor diagonal below the main one and
0 everywhere else,
except on the first row where will stand $n_i$ different $W$ fields.
In the last $p\times p$ matrix will
stand $(p^2-1)$ different $W$ fields, as well as on the
diagonal $p$-times the same
$W$-field. This last field depends on the others $k$ fields on the
main diagonal, since the trace of
an $S\ell(n)$ matrix is zero. Note that the fields relative to the
commutant of \ch, constituted by
$G\ell(p)$ and the $(k-1)$ $U(1)$, generate the
Kac-Moody part (spin-one fields) of
the \cw(\cg, \ch) algebra.

\indent

Now, let us decompose the rest of the $S\ell(n)$ matrix into
$n_i\times n_j$ $(i\neq j)$, $n_i\times p$ and $p\times n_i$
submatrices. In each matrix of this type, above the block diagonal
part, the $W$
fields will show up only on the first row, the rest of the matrix
entries being 0. There will be $p$ different fields for a matrix
$n_i\times p$ or $p\times n_i$.
For a $n_i\times n_j$ matrix, the number of different fields will
be $min(n_i,n_j)$; when $n_j\leq n_i$ the first
row will then be filled by these
$n_j$ fields, but when $n_j>n_i$, the first row will be
made with $n_j-n_i$ zeros followed by the $W$-fields in position
$n_j-n_i+1,\ n_j-n_i+2,\ \dots,\ n_j$. Below the diagonal
part, one uses the same submatrix decomposition.
The number of different $W$
fields in each case will be the same as above, except that
now they will occupy
the last column with again $p$ fields for the matrix $p\times n_i$ or
$n_i\times p$ and for the $n_i\times n_j$ matrix, $min(n_j,n_i)$ fields
put in the first, second, up to $min(n_j,n_i)$-th row.

In the case \ch=\cg, the $(n-1)$ different $W$ fields of the Abelian
\cw(\cg, \cg) algebra will occupy the second, third, up to $n$-th
entries of the first row\footnote{In this special case, there is {\it
a priori} one field on the diagonal (on the first row) but it is set
to zero by the traceless condition.}, the
rest of the matrix being made of 0 and 1: then we recognize the so-called
horizontal gauge.
By extension, we will call the gauge described above
the Generalized Horizontal
Gauge (G.H.G.).

We also define the Ordered Generalized Horizontal Gauge (O.G.H.G.) by
disposing in the $S\ell(n)$
matrix the different $S\ell(n_i)$ of \ch\ in dicreasing order,
i.e. $S\ell(n_i)$ will show up before $S\ell(n_j)$ if $n_i>n_j$. This
gauge is such that the first row (or column) of each submatrix is
completely filled with $W$ fields (no zero: see for instance the matrix
\ref{eq.sl11}).

\indent

Let us illustrate this construction in a particular case:
$\cg=S\ell(11)$ and $\ch=S\ell(3)\oplus S\ell(2)\oplus
S\ell(2)$. Then $S\ell(p)=S\ell(4)$.
One can compute that the corresponding \cw-algebra
will be generated by 58 fields, the Kac-Moody part being
made by $S\ell(4)$ and three $U(1)$
(We recall that the total $S\ell(11)$ matrix is traceless).
In the matrix just below, we have explicitly written the entries that
correspond to a 1 or a $W$ field: the rest of the matrix is filled by
zeros. The positions of
the $W$-fields is specified by a star ($\ast$).

\indent

\be
\left(\begin{array}{ccc|cc|cc|cccc}
* & * & * & * & * & * & * & * & * & * & *\\
1 & {} & {} & {} & {} & {} & {} & {} & {} & {} & {}\\
{} & 1 & {} & {} & {} & {} & {} & {} & {} & {} & {}\\
\hline
{} & {} & * & * & * & * & * & * & * & * & *\\
{} & {} & * & 1 & {} & {} & {} & {} & {} & {} & {}\\
\hline
{} & {} & * & {} & * & * & * & * & * & * & *\\
{} & {} & * & {} & * & 1 & {} & {} & {} & {} & {}\\
\hline
{} & {} & * & {} & * & {} & * & * & * & * & *\\
{} & {} & * & {} & * & {} & * & * & * & * & *\\
{} & {} & * & {} & * & {} & * & * & * & * & *\\
{} & {} & * & {} & * & {} & * & * & * & * & *
\end{array}\right)
\label{eq.sl11}
\ee

\indent

In the matrix (\ref{eq.sl11}), the horizontal and vertical lines are
drawn to make apparent the submatrices $n_i\times p$, $p\times n_i$ and
$n_i\times n_j$ described above. Note that this matrix corresponds to
an O.G.H.G.: as a consequence, the lines of "stars" are "full".

\indent

Now it is rather obvious that the constraints associated
to the O.G.H.G. of the couple (\cg, \ch) will be included into the
constraints associated to
the G.H.G. of (\cg, \ch$'$) when $\ch\subset\ch'$.
Thus, with the help of property \ref{prop1}, we conclude that we can
gauge \cw(\cg,\ch) to get \cw(\cg,\ch$'$).
Note that although we have
started from an ordered G.H.G, after reduction, the new gauge may be not
ordered anymore.

Let us take the opportunity of matrix (\ref{eq.sl11})
to treat explicitly an example. We choose,
for example, to reduce the given algebra \cw(\cg, \ch)  with
$\cg=S\ell(11)$ and $\ch=S\ell(3)\oplus S\ell(2)\oplus
S\ell(2)$ up to \cw(\cg, \ch$'$) with $\ch'=S\ell(3)\oplus S\ell(4)$.
Then the $W$-fields in position (1,4), (6,5), (6,6), $\dots$,
(6,11) will disappear as
well as those in position (7,3), (7,5) and also those in
position (8,5), (9,5), (10,5), (11,5).
With the same conventions as in
(\ref{eq.sl11}), the gauged matrix becomes:

\be
\left(\begin{array}{ccc|cccc|cccc}
* & * & * & {} & * & * & * & * & * & * & *\\
1 & {} & {} & {} & {} & {} & {} & {} & {} & {} & {}\\
{} & 1 & {} & {} & {} & {} & {} & {} & {} & {} & {}\\
\hline
{} & {} & * & * & * & * & * & * & * & * & *\\
{} & {} & * & 1 & {} & {} & {} & {} & {} & {} & {}\\
{} & {} & * & {} & 1 & {} & {} & {} & {} & {} & {}\\
{} & {} & {} & {} & {} & 1 & {} & {} & {} & {} & {}\\
\hline
{} & {} & * & {} & {} & {} & * & * & * & * & *\\
{} & {} & * & {} & {} & {} & * & * & * & * & *\\
{} & {} & * & {} & {} & {} & * & * & * & * & *\\
{} & {} & * & {} & {} & {} & * & * & * & * & *
\end{array}\right)
\ee

\indent

As announced, the new gauge is a non-ordered G.H.G., and some lines of
"stars" are "half-empty". The inclusion of the
constraints appears naturally, and one can check
independently that we have the right number of \cw(\cg, \ch$'$) fields,
i.e. 44.

\sect{An example: reduction of $\cw(S\ell(4),\ch)$-algebras.\label{sec5}}

\indent

As an illustrative example of the developed method, let us consider the Lie
algebra $S\ell(4)$, which leads to four different $\cw$-algebras by Hamiltonian
reduction, namely $\cw [S\ell(4),S\ell(2)]$, $\cw [S\ell(4),2 S\ell(2)]$,
$\cw [S\ell(4),S\ell(3)]$ (non Abelian cases) and $\cw [S\ell(4)]$
(Abelian case). Among them, the model corresponding to
$\cw [S\ell(4),S\ell(2)]$ is the less constrained one. Thus, let us start
with this algebra. Using classical arguments, we directly
deduce that a basis of $W$ fields is given by one spin 2, four spin 1
and four spin 3/2 primary fields. Following section \ref{sec4}, one chooses
the generalized horizontal gauge:

\be
J =
\left(\begin{array}{cc|cc}
2 W_1 & W_2 & W_{3/2}^{+-} & W_{3/2}^{++} \\
1 & 0 & 0 & 0 \\ \hline
0 & W_{3/2}^{-+} & - W_1 + W_1^0 & W_1^+ \\
0 & W_{3/2}^{--} & W_1^- & - W_1 - W_1^0
\end{array}\right)
\ee

Notice that in this gauge the fields are not primary. The KM part
of the $\cw [S\ell(4),S\ell(2)]$ algebra is generated by the four spin
one fields which form the algebra $S\ell(2)^{(1)} \oplus U(1)^{(1)}$.

{}From the property \ref{prop1}
and the choice of the gauge, we deduce that one can go
from the $\cw [S\ell(4),S\ell(2)]$ algebra to the three others by a
suitable set of constraints. More precisely, setting as (first class)
constraint the condition $W_1^- = 1$, the associated gauge transformation
\be
W(x) \longrightarrow \wh{W}(x) = W(x) + \int dz\
\{\alpha(z)W_1^-(z),W(x)\} + ...
\ee
leads to the following transformed fields:
\be
\begin{array}{ll}
\wh{W}_1 = W_1 & \hs{10} \wh{W}_1^0 = W_1^0 + \alpha \\
\wh{W}_1^+ = W_1^+ - 2\alpha W_1^0 + \prt \alpha - \alpha^2 &
\hs{10} \wh{W}_2 = W_2 \\
\wh{W}_{3/2}^{+-} = W_{3/2}^{+-} & \hs{10}
\wh{W}_{3/2}^{++} = W_{3/2}^{++} - \alpha W_{3/2}^{+-} \\
\wh{W}_{3/2}^{--} = W_{3/2}^{--} & \hs{10}
\wh{W}_{3/2}^{-+} = W_{3/2}^{-+} + \alpha W_{3/2}^{--} \\
\end{array}
\ee
The gauge fixing is determined by imposing
$\wh{W}_1^0 = -W_1$ i.e. $\alpha = -W_1^0-W_1$.
The gauge fixed fields will read
\be
\begin{array}{ll}
\wh{W}_1 = W_1 & \hs{10} \wh{W}_1^+ = W_1^+ + W_1^0.W_1^0
-W_1.W_1 - \prt W_1^0 -\prt W_1\\
\wh{W}_{3/2}^{+-} = W_{3/2}^{+-} & \hs{10}
\wh{W}_{3/2}^{++} = W_{3/2}^{++} + W_1^0.W_{3/2}^{+-}
+W_1.W_{3/2}^{+-}\\
\wh{W}_{3/2}^{--} = W_{3/2}^{--} & \hs{10}
\wh{W}_{3/2}^{-+} = W_{3/2}^{-+} - W_1^0 W_{3/2}^{--} -
W_1.W_{3/2}^{--}\\
\wh{W}_2 = W_2 & \\
\end{array}
\label{hatf}
\ee
The corresponding current matrix $\wh{J}$ is then

\be
\wh{J} =
\left(\begin{array}{cc|cc}
2 \wh{W}_1 & \wh{W}_2 & \wh{W}_{3/2}^{+-} & \wh{W}_{3/2}^{++} \\
1 & 0 & 0 & 0 \\ \hline
0 & \wh{W}_{3/2}^{-+} & -2\wh{W}_1 & \wh{W}_1^+ \\
0 & \wh{W}_{3/2}^{--} & 1 & 0
\end{array}\right)
\ee

The commutation relations among the $\wh{W}$ fields can be computed
explicitly using the relations (\ref{hatf}) and the commutation relations of
$\cw [S\ell(4),S\ell(2)]$. One finds (and this can be checked also with
the form of the $\wh{J}$ current) that the $\wh{W}$ fields generate the
algebra $\cw [S\ell(4),2 S\ell(2)]$, with four spin 2 and three spin 1 fields.
The primary fields are given by
\bea
&\mbox{stress-energy tensor}& \wh{W}_2 + \wh{W}_1^+ + 4\wh{W}_1.\wh{W}_1
\nonumber \\
&\mbox{spin 1 fields}& \wh{W}_1, \hs{5} \wh{W}_{3/2}^{+-},
\hs{5} \wh{W}_{3/2}^{--} \nonumber \\
&\mbox{spin 2 fields}& \wh{W}_1^+ - \wh{W}_2 + 2\prt \wh{W}_1, \hs{5}
\wh{W}_{3/2}^{++} - \frac{1}{2} \prt \wh{W}_{3/2}^{+-},
\hs{5} \wh{W}_{3/2}^{-+} + \frac{1}{2} \prt \wh{W}_{3/2}^{--}
\ena

Now, one can perform a secondary reduction on the algebra $\cw [S\ell(4),
2 S\ell(2)]$ with the choice of (first class) constraints
$\wh{W}_{3/2}^{-+} = 1$ and $\wh{W}_{3/2}^{--} = 0$,
with the associated gauge transformation
\be
\wh{W}(x) \longrightarrow \wt{W}(x) = \wh{W}(x)
+ \int dz \{(\alpha \wh{W}_{3/2}^{-+} + \beta \wh{W}_{3/2}^{--})(z),
\wh{W}(x)\} + ...
\ee
The transformed fields are:
\bea
\wt{W}_1 &=& \wh{W}_1 + \alpha/2 \nonumber \\
\wt{W}_1^+ &=& \wh{W}_1^+ - \beta \nonumber \\
\wt{W}_2 &=& \wh{W}_2 - 4 \alpha \wh{W}_1 + 2 \prt \alpha + \beta
- \alpha^2
\ena
and more complicated expressions for $\wt{W}_{3/2}^{+-}$ and
$\wt{W}_{3/2}^{++}$. The gauge fixing is obtained by imposing
$\wt{W}_1 = 0$ and $\wt{W}_1^+ = 0$ that is
$\alpha = - 2\wh{W}_1$ and $\beta = \wh{W}_1^+$.
The corresponding current matrix $\wt{J}$ takes the form

\be
\wt{J} =
\left(\begin{array}{cccc}
0 & \wt{W}_2 & \wt{W}_{3/2}^{+-} & \wt{W}_{3/2}^{++} \\
1 & 0 & 0 & 0 \\
0 & 1 & 0 & 0 \\
0 & 0 & 1 & 0
\end{array}\right)
\label{hatt}
\ee

The obtained gauged $\cw$-algebra is in this case the algebra
$\cw [S\ell(4)]$.

\indent

Finally, we could have done the reduction from $\cw [S\ell(4),S\ell(2)]$
to $\cw [S\ell(4)]$ directly by taking the constraints
$W_1^- = 1$, $W_{3/2}^{-+} = 1$ and $W_{3/2}^{--} = 0$
which generate the gauge transformation
\be
W(x) \longrightarrow \wt{W}(x) = W(x) + \int dz \
\{(\alpha W_{3/2}^{-+} + \beta W_{3/2}^{--} + \gamma W_1^-)(z),W(x)\} + ...
\ee
The transformed fields are now
\be
\begin{array}{l}
\wt{W}_1 = W_1 + \alpha/2 \\
\wt{W}_1^0 = W_1^0 - \alpha/2 + \gamma \\
\wt{W}_1^+ = W_1^+ - \beta - 2\gamma W_1^0 + \prt \gamma - \gamma^2 +
\alpha \gamma/2 \\
\wt{W}_2 = W_2 + 2 \prt \alpha + \alpha W_1^0 - 3 \alpha W_1 + \beta
+ \alpha \gamma /2 - \alpha^2
\end{array}
\ee
(the expressions for $\wt{W}_{3/2}^{+-}$ and $\wt{W}_{3/2}^{++}$ are
rather heavy). We choose as gauge fixing the conditions
$\wt{W}_1 = \wt{W}_1^0 = \wt{W}_1^+ = 0$
which lead to
$\alpha = -2W_1, \beta = W_1^+ + W_1.W_1^0 + W_1^0.W_1^0 - \prt W_1
- \prt W_1^0, \gamma = - W_1 - W_1^0$,
and the gauge fixed fields $\wt{W}_2$, $\wt{W}_{3/2}^{+-}$ and
$\wt{W}_{3/2}^{++}$ generate the algebra $\cw [S\ell(4)]$
with the current matrix (\ref{hatt}).

The other cases can be treated similarly. We summarize in the table
\ref{tab.sl4}
the complete secondary reduction scheme in the case of the $\cw$-algebras
associated to $S\ell(4)$.

\begin{table}[p]
\[
\begin{array}{c}
\left(\begin{array}{cc|cc}
2 W_1 & W_2 & W_{3/2}^{+-} & W_{3/2}^{++} \\
1 & 0 & 0 & 0 \\ \hline
0 & W_{3/2}^{-+} & -W_1 + W_1^0 & W_1^+ \\
0 & W_{3/2}^{--} & W_1^- & -W_1 - W_1^0
\end{array}\right)
\\
\\
{\bf \cw [S\ell(4),S\ell(2)]}
\\
\begin{array}{ccc}
&& \\
\hs{10}W_1^- = 1 \hs{10} \swarrow &&\hs{-14} \searrow
\hs{10} W_{3/2}^{-+} = 1, W_{3/2}^{--} = 0
\\ && \\
\left(\begin{array}{cc|cc}
2 \wh{W}_1 & \wh{W}_2 & \wh{W}_{3/2}^{-+} & \wh{W}_{3/2}^{--} \\
1 & 0 & 0 & 0 \\ \hline
0 & \wh{W}_{3/2}^{-+} & -2\wh{W}_1 & \wh{W}_1^+ \\
0 & \wh{W}_{3/2}^{--} & 1 & 0
\end{array}\right)
&
\begin{array}{c} | \\ | \\ \\ W_{3/2}^{-+} = 1, W_{3/2}^{--} = 0, W_1^- = 1
\\ \\ | \\ \downarrow \end{array}
&
\left(\begin{array}{ccc|c}
\wh{W}_1 & \wh{W}_2 & \wh{W}_{3/2}^{+-} & \wh{W}_{3/2}^{++} \\
1 & 0 & 0 & 0 \\
0 & 1 & 0 & 0 \\ \hline
0 & 0 & \wh{W}_1^- & - \wh{W}_1
\end{array}\right)
\\
{\bf \cw [S\ell(4),2 S\ell(2)]} & \hs{50} & {\bf \cw [S\ell(4),S\ell(3)]}
\\
&& \\
\wh{W}_{3/2}^{+-} = 1, \wh{W}_{3/2}^{++} = 0 \hs{10} \searrow &&
\hs{-24}\swarrow \hs{10} \wh{W}_1^- = 1
\end{array}
\\
\\
\left(\begin{array}{cccc}
0 & \wt{W}_2 & \wt{W}_{3/2}^{+-} & \wt{W}_{3/2}^{++} \\
1 & 0 & 0 & 0 \\
0 & 1 & 0 & 0 \\
0 & 0 & 1 & 0
\end{array}\right)
\\
\\
{\bf \cw [S\ell(4)]}
\end{array}
\]
\caption{Secondary reductions for $\cw[S\ell(4), \ch]$ algebras}
\label{tab.sl4}
\end{table}
The table \ref{tab.sl4} is actually completely general.
The algebra $\cw [S\ell(n),S\ell(2)]$
corresponds always to the less constrained model (i.e. those which contains the
most number of fields), while the Abelian $\cw [S\ell(n)]$ algebra corresponds
to the most constrained model (which contains exactly $n-1$ fields). From the
$\cw [S\ell(n),S\ell(2)]$ algebra, one can obtain all the other algebras.

\sect{Conclusion}

We have shown that $W$-currents can be constrained to provide a new
\cw-algebra in
the same way that conditions on Kac-Moody currents lead to a \cw-algebra.

Two \cw-algebras can be connected by this mechanism as soon as the set of first
and second class constraints of the first algebra is included in the set of
constraints of the second one. For $\cw[S\ell(n),\ch]$ algebras
the existence of a suitable
gauge, denoted Generalized Horizontal Gauge, insures that
\cw(\cg, \ch) can be
``gauged'' to \cw(\cg, \ch$'$) as soon as $\ch\subset\ch'$. Chains of
\cw-algebras
can then be constructed
in relation with the chains of regular, semi-simple subalgebras in $S\ell(n)$.
Although
different, there exists similarity between these chains of subalgebras
embeddings and the classification of orbits in the adjoint representation
of \cg:
could remarkable geometrical properties known for the
\cg-orbits \cite{Michel} be of some relevance
for our problem?
We have not yet been able to give a pertinent answer to this
question.

Let us stress that the above study has been done only for classical
\cw-algebras. The
quantum case needs to be considered. It would also be of some interest to
generalize the above results to all series of simple algebras and
superalgebras.

Finally, it looks to us specially worthwhile
to develop the \cw-gauge aspects of
this approach. In this respect, the Lagrangian formalism for such a \cw-gauge
theory is under study and will soon be presented.

\end{document}